\documentclass[aps,prb,superscriptaddress,twocolumn]{revtex4-1}

\usepackage{graphics}
\usepackage{graphicx} 
\usepackage{dcolumn}
\usepackage{bm}
\usepackage{epstopdf}
\usepackage[usenames, dvipsnames]{color}

\begin{document} 



\title{Magnetic excitations in the quasi-2D ferromagnet Fe$_{\rm 3-x}$GeTe$_2$ measured with inelastic neutron scattering}



\author{S.~Calder}
\email{caldersa@ornl.gov}
\affiliation{Neutron Scattering Division, Oak Ridge National Laboratory, Oak Ridge, Tennessee 37831, USA.}

\author{A.~I.~Kolesnikov}
\affiliation{Neutron Scattering Division, Oak Ridge National Laboratory, Oak Ridge, Tennessee 37831, USA.}

\author{A.~F.~May}
\affiliation{Materials Science and Technology Division, Oak Ridge National Laboratory, Oak Ridge, Tennessee 37831, USA}


\begin{abstract}	
Fe$_{\rm 3-x}$GeTe$_2$ is an itinerant ferromagnet composed of two-dimensional layers weakly connected by van der Waals bonding that shows a variety of intriguing phenomena. Inelastic neutron scattering measurements on bulk single crystals of Fe$_{2.75}$GeTe$_2$ were performed to quantify the magnetic exchange interaction energies and anisotropy. The observed inelastic excitations are indicative of dominant in-plane correlations with negligible magnetic interactions between the layers. A spin-gap of 3.9 meV is observed allowing a  measure of the magnetic anisotropy. As the excitations disperse to their maximum energy ($\sim$65 meV) they become highly damped, reflective of both the magnetic site occupancy reduction of 25{\%} on one Fe sublattice and the itinerant interactions. A minimal model is employed to describe the excitation spectra and extract nearest neighbor magnetic exchange interaction values. The temperature evolution of the excitations are probed and correlations shown to persist above T$\rm _c$, indicative of low dimensional magnetism. 
\end{abstract}


\maketitle

\section{\label{sec:Introduction}Introduction}

Reducing the dimensionality of a compound to topologically constrained layers can create fundamental phenomena beyond well-established classical behavior. In this context graphene, formed from the isolation of van der Waals (VDW) bonded two-dimensional (2D) layers from graphite by exfoliation to a single honeycomb monolayer, ignited widespread interest \cite{NatureGraphene}. Exotic quantum relativistic phenomena, such as Dirac semi-metals and quantum anomalous Hall insulators, have been predicted in graphene and related materials ranging from isolated 2D monolayers to quasi-2D bulk materials with VDW bonded layers \cite{ZhangQHallGraphene, Yu61}. Particular interest has extended to VDW layered materials beyond graphene that contain magnetic ions \cite{McGuirehalides2017,Burch2018}. This is driven by the potential for intriguing quasi-low dimensional quantum phenomena related to the spin degree of freedom and future applicability in novel spintronic devices based on these paradigms. As with graphene a promising route to achieve low dimensional behavior is to start with suitable bulk compounds with VDW bonded layers and then create isolated 2D layers in a top-down approach. 

Investigating the bulk qausi-2D VDW compound is often the first step and can reveal a plethora of intriguing phenomena due to the quasi-2D isolated magnetic layers. In addition bulk materials are amenable to a variety of probes not well suited to monolayers due to requirements for large mass to achieve observable signals and suitable statistics. In this respect inelastic neutron scattering (INS), which typically requires gram sized samples, is particularly powerful. INS allows the quantitative extraction of the spin Hamiltonian, which contains information on the magnetic ordering, exchange interactions and anisotropy. Such studies on bulk crystals can reveal empirical information that can act as a bridge to understand and predict single layer behavior. Indeed INS investigations on bulk magnetic qausi-2D VDW materials have proven fruitful with studies on CrSiTe$_3$ \cite{PhysRevB.92.144404} and the family of $M$PS$_3$, with $M$=Fe \cite{0953-8984-24-41-416004,PhysRevB.94.214407}, $M$=Mn \cite{0953-8984-10-28-020} and $M$=Ni \cite{PhysRevB.98.134414}.

In the context of magnetic VDW materials Fe$_{\rm 3-x}$GeTe$_2$ (FGT) is of current interest. Studies of the bulk material have shown a non-trivial anomalous Hall effect \cite{PhysRevB.96.134428,TopologicalHallFGT,PhysRevB.97.165415}, interesting electronic properties \cite{Zhangeaao6791}, strong electron correlations \cite{PhysRevB.93.144404} and unusual bubble and stripy magnetic domain structures \cite{PhysRevB.97.014425,doi:10.1063/1.4961592}. The magnetism is strongly anisotropic with a large magnetic anisotropy energy (MAE), required for storage materials \cite{verchenko_inorgchem, doi:10.1063/1.4961592, PhysRevB.93.014411, PhysRevB.93.134407, PhysRevB.93.144404}. These results have led to investigations considering the behavior of FGT as the number of layers are reduced from bulk to approach a monolayer. Theoretically the monolayers were predicted to be stable with formation energy $\Delta \rm E_F$, defined as the energy difference between monolayer and bulk of $\Delta \rm E_F$=50meV/atom, well below the upper bound of $\Delta \rm E_F$=200meV/atom considered stable to create monolayers. Additionally the single layer phonon dispersions are predicted to be stable \cite{PhysRevB.93.134407}. Recent experimental reports have shown isolation of atomically thin FGT from the bulk compound \cite{MayNatureMaterials}, with an intriguing study showing ionic liquid gating can increase T$\rm _c$ to room temperature \cite{GateTuneableFGT}. The persistence of itinerant ferromagnetism, large coercivity, strong out-of-plane anisotropy and the anomalous hall effect have all been shown in reduced layered materials, with a cross-over from 3D-2D Ising magnetism \cite{GateTuneableFGT, FGT_ncomm, MayNatureMaterials,ISIwaferscale}. Collectively these properties make FGT a promising candidate for hetorostructure-based spintronic applications.  

\begin{figure}[tb]
	\centering         
	\includegraphics[trim=0cm 3cm 0cm 0cm,clip=true, width=1.0\columnwidth]{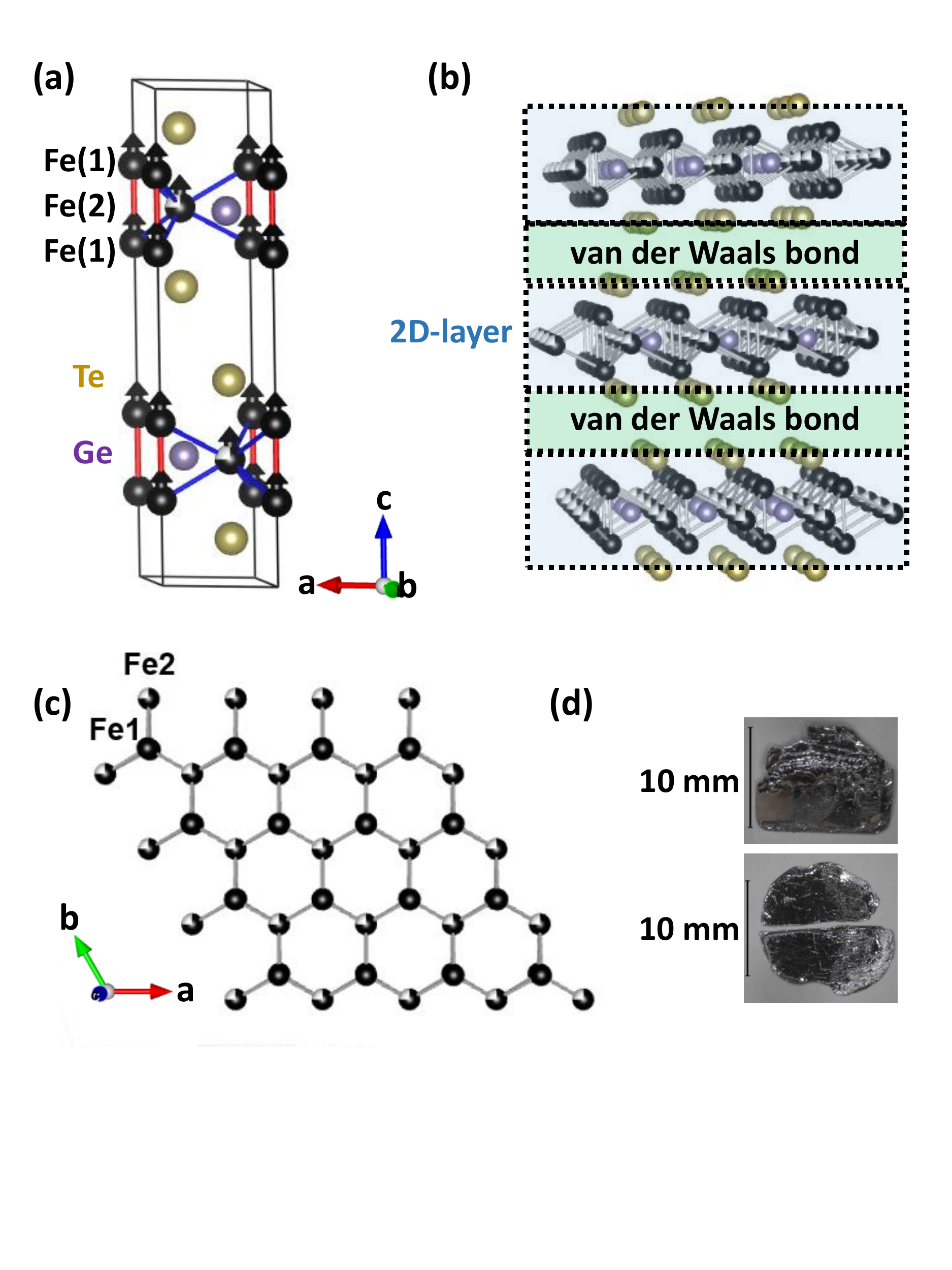}           
	\caption{\label{Fig1} (a) Structural/magnetic unit cell of Fe$_{\rm 3-x}$GeTe$_2$. Arrows indicate the magnetic moments on the Fe ions. The nearest neighbor J$_1$-J$_1$ (red) and next nearest nighbor J$_1$-J$_2$ (blue) interactions are shown. (b) The qausi-2D layered structure is highlighted. (c) Top-down view of the hexagonally arranged magnetic Fe ions in the ab-plane. (d) Fe$_{2.75}$GeTe$_2$ flux grown single crystals.}
\end{figure}

FGT crystallizes in the hexagonal space group $P6_3/mmc$ with 2D layers of Fe$_{\rm 3-x}$Ge sandwiched between nets of Te ions that are weakly connected by VDW bonding \cite{DeiserothFGT}, see Fig.~\ref{Fig1}. The Fe ions are located on two inequivalent sites, Fe(1) at (0,0,z) and Fe(2) at ($\frac{1}{3},\frac{2}{3},\frac{1}{4}$). The Fe ions form a hexagonal motif in the ab-plane and order ferromagnetically with a high T$\rm _c$ in the range T$\rm _c$$\approx$150-230 K. The wide temperature range is driven by alteration of the Fe site occupancy, that occurs only on the Fe(2) site, altering the lattice constants through chemical pressure and thereby tuning the magnetic interaction and anisotropy energies \cite{PhysRevB.93.014411}. A reduction in Fe(2) occupancy leads to a reduction in $\rm T_c$ and anisotropy.  While magnetization measurements are indicative of ferromagnetic ordering an anomaly in low field warming measurements within the ordered phase is suggestive of a potential crossover to more exotic magnetism \cite{2053-1583-4-1-011005}. The itinerant nature of the magnetism in FGT is consistent experimentally with reduced ordered moments measured with neutron diffraction \cite{PhysRevB.93.014411, verchenko_inorgchem}.


Here we present a single crystal INS investigation of FGT with 25{\%} Fe(2) vacancies to probe the collective magnetic excitations. The results reveal indications for the low dimensionality of the magnetic correlations in the bulk material and provide experimental values for the magnetic exchange interactions and magnetic anisotropy through the use of a minimal model that captures the essential features of the excitations.

\section{\label{sec:Expt_Theory}Experimental and theoretical details}

\subsection{Sample preparation}

The single crystals for this study were prepared by the flux method and characterized as described in Ref.~\onlinecite{PhysRevB.93.014411}. They show a ferromagnetic transition at T$\rm _c$=150 K. These crystals grow with an Fe(2) content of 0.75, with the Fe(1) site being fully occupied. This preparation method produces single crystals of over 0.6 grams and 10$\times$10 mm$^2$. This is significantly larger than those reported from growth with vapor transport, making the attainment of the required gram sized sample feasible for INS. Six single crystals were coalgined using the CG-1B neutron alignment station at the High Flux Isotope Reactor (HFIR), Oak Ridge National Laboratory (ORNL), to achieve a total mass of 1.95 g. To avoid straining the sample or introducing additional background adhesive was not used, instead the crystal mount only consisted of Al and the samples. The chosen plane was [H0L] for all discussed measurements with a FWHM of 1.2$^\circ$ for the array based on rocking scans on CG-1B.

\subsection{Inelastic neutron scattering}

INS measurements were performed on the SEQUOIA \cite{1742-6596-251-1-012058, doi:10.1063/1.4870050}  and  ARCS \cite{doi:10.1063/1.3680104, doi:10.1063/1.4870050}  time-of-flight spectrometers at the Spallation Neutron Source (SNS), ORNL. The measurements on ARCS were focused on measuring the full energy range of the spectrum with suitably chosen higher incident energies (E$\rm _i$) and subsequently coarser resolution. Energies measured were E$\rm _i$=120 meV, with Fermi chopper at 600 Hz and T$_0$=120 Hz, and E$\rm _i$=50 meV, with Fermi chopper at 360 Hz and T$_0$=60 Hz. All measurements were performed at base temperature of a closed-cycle refrigerator (CCR) of 10 K. The measurements on SEQUOIA utilized the finer resolution available on the instrument from the longer neutron flight path and chopper package to resolve the low energy scattering and determine any spin-gap energy and access temperature dependence. Energies measured were E$\rm _i$=10 meV, Fermi chopper 180 Hz and T$_0$=30 Hz, and E$\rm _i$=30 meV, Fermi chopper 120 Hz and T$_0$=30 Hz. Measurements were performed at 10 K, 160 K and 295 K. On both INS instruments measurements were performed over a range of crystal rotations in small step sizes of a few degrees. The data were reduced using Mantid \cite{ARNOLD2014156} for each rotation step and combined together for the particular E$\rm _i$ with the Horace software \cite{EWINGS2016132} to produce a four-dimensional (H,K,L,E) reciprocal and energy space data set. An empty sample holder was measured under the same conditions as the sample to allow for a background subtraction. Simulations of the scattering were performed with SpinW \cite{spinW}.

\section{\label{sec:Res_Discuss}Results and Discussion}

\begin{figure}[tb]
	\centering         
	\includegraphics[trim=0cm 1.8cm 0cm 1cm,clip=true, width=1.0\columnwidth]{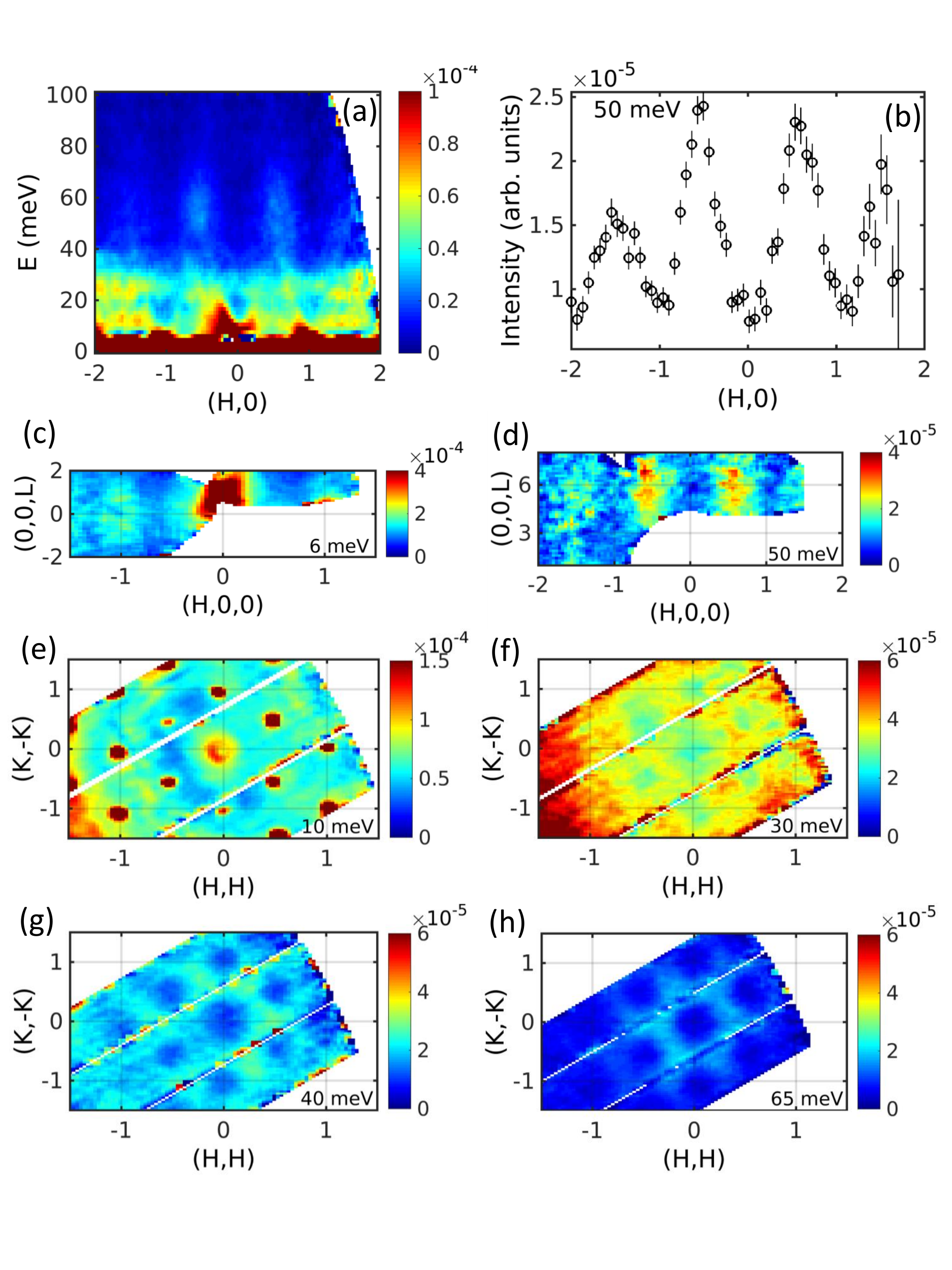}           
	\caption{\label{Fig2} (a) INS measurements with E$\rm _i$=120meV that captured the full range of observable magnetic excitations. Slice covers $-0.1$$\leq$K$\leq$$0.1$ and $10$$\leq$L$\leq$$10$ r.l.u. (b) Constant energy cut at 50 meV, with an energy range of $\pm$5meV. Cut covers $-0.1$$\leq$K$\leq$$0.1$ and $10$$\leq$L$\leq$$10$ r.l.u. (c) Inelastic scattering in the (H0L) plane at 6 meV, $\pm$2meV and -0.1$\leq$K$\leq$0.1 r.l.u., and (d) 50 meV, $\pm$5meV and $-0.15$$\leq$K$\leq$$0.15$ r.l.u. (e)-(g) Inelastic scattering in the (HH,K-K) plane at constant energies of 10 meV, 30 meV, 40 meV and 65 meV, with an energy range of $\pm$5meV. All measurements were performed at 10 K.}
\end{figure}

We begin by considering the INS measurements performed with the highest incident energy of E$\rm _i$=120 meV. Inelastic excitations were observed to extend up to 65 meV, see Fig.~\ref{Fig2}(a)-(b), with the expected increased intensity at low Q of magnetic scattering. Projecting the data onto the (H0L) plane at select inelastic energy ranges, Fig.~\ref{Fig2}(c)-(d), produced rod-like scattering with in-plane H momentum dependence but negligible out of plane L momentum dependence. This provides indications of the 2D nature of the magnetic exchange interactions in bulk FGT, however, we note there may be additional influence from the reduced Fe(2) occupation causing a loss of coherence along L due to stacking faults. In any case, the lack of L dependence allowed the data to be integrated over a wide L range and therefore access a wider range of reciprocal and energy space than would otherwise be possible. The data was checked to confirm this procedure produced identical results to that carried out over a limited L range. This method can be attributed, however, to the contributions of increased intensity in the range 0-30meV, which is attricbuted to phonon scattering. Constant energy cuts are shown in Fig.~\ref{Fig2}(e)-(f) for the (HH,K-K) in-plane scattering. Well-defined momentum dependence is observed, although the scattering is broader than instrument resolution. As expected for spin wave excitations the scattering originates from the magnetic Brillouin zone center and disperses out to the zone boundary.

\begin{figure}[tb]
	\centering         
	\includegraphics[trim=0cm 4cm 0cm 4cm,clip=true, width=1.0\columnwidth]{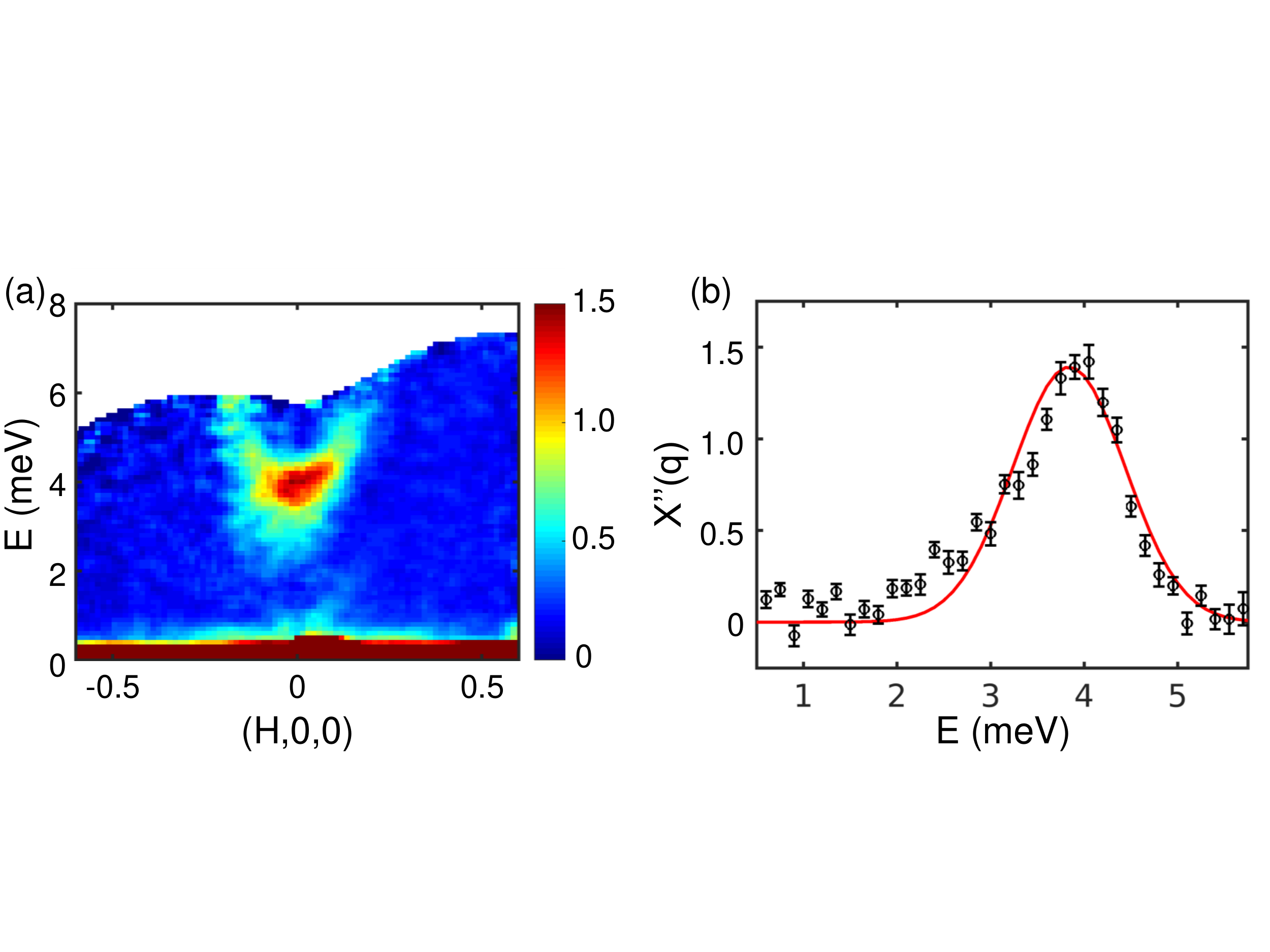}           
	\caption{\label{Fig3} (a) INS measurements with E$\rm _i$=10 meV around the magnetic zone center. (b) Cut through the magnetic zone center of (0,0,0) at low energy (circles) compared to model calculations (line) to define the spin-gap energy of 3.9(1) meV.}
\end{figure}

Measurements with low incident energy were performed to define the presence of a spin gap at the Brillouin zone (BZ) center that indicates the strength of the magnetic anisotropy. Results with the lowest $\rm E_i$=10 meV are shown in Fig.~\ref{Fig3} that reveal a spin-gap. Temperature dependence was measured and the data Bose corrected to ensure the scattering observed is not contaminated by phonon scattering. 

The data therefore is characterized by well-defined $\rm S(q,\omega)$ magnetic scattering at low energy that broaden out as they disperse up to their maximum energy of 65 meV. The broad nature of the magnetic excitations likely has two principle causes. Primarily, the sample contains 25$\%$ vacancies on the Fe(2) site that introduces strong disorder and will strongly damp the magnetic excitations at zone boundary. This is indeed what is observed and could be further confirmed by performing INS on non-deficient FGT samples if suitable mass could be obtained. Another contributing factor is the itinerant nature of the electrons in FGT which results in intrinsically broad signals as the inelastic energy transfer increases, as observed in several systems including Fe-based superconductors \cite{PhysRevLett.102.187206} and nickelates \cite{LaNiO3itinerant}. The consequence of the broad features means that the dispersion is not well defined and this provides limits to modeling the data and the number of exchange terms that can be uniquely defined. As a result we focus on providing a quantitative minimal model that describes the excitations. To achieve this we utilize a spin wave approach by invoking an effective two-dimensional local moment Hamiltonian with nearest neighbor Fe(1)-Fe(1) ($J_1$) and next nearest neighbor Fe(1)-Fe(2) ($J_2$) exchange interactions with a single-ion anisotropic term to describe the spin-gap:

\begin{equation}
\label{ham}
\mathcal{H}=\sum_{i,j}J_{ij} \mathbf{S}_i\cdot\mathbf{S}_j +  \sum_{i,z} -D_{z} (S_i^{z})^2
\end{equation}

This approach is an approximation and neglects the itinerant nature of the magnetism. However it has been applied extensively in other studies that diverge from ideal local magnetism to extract robust empirical parameters, for example in the Fe-based superconductors \cite{CaFe2As2Dai, RbFeSeDai}. Moreover, utilizing a similar Heisenberg Hamiltonian to Equation (1) has been shown to be applicable to FGT in Ref.~\onlinecite{GateTuneableFGT}. Extensions beyond this model would be of interest for future studies, particularly those involving FGT with a fully occupied Fe(2) site that should remove damping from site disorder.  

\begin{figure}[tb]
	\centering         
	\includegraphics[trim=0cm 2.5cm 0cm 5cm,clip=true, width=1.0\columnwidth]{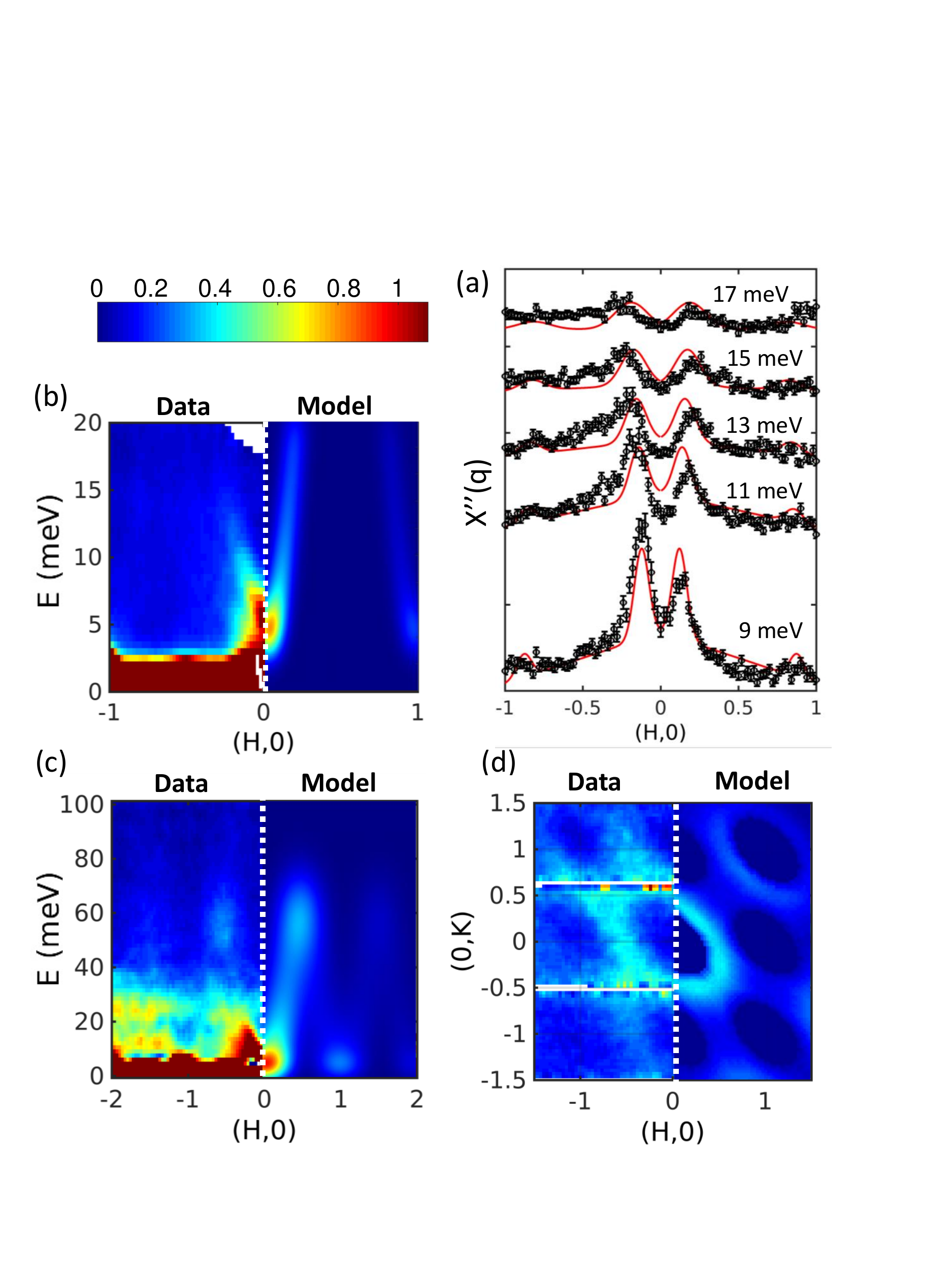}           
	\caption{\label{Fig4} Comparison of INS data with a spin wave model. (a) Constant energy data with E$\rm_i$=50 meV (circles) compared to the model (lines). (b) Energy dispersion from E$\rm_i$=50 meV and (c) E$\rm_i$=120 meV compared to model simulations. (d) Data and model of a constant energy slice of 60 meV in the (H,K) plane.}
\end{figure}

As a suitable starting point we begin by modeling the data with $J_1$=$J_2$ given the similar bond distance of $J_1$ and $J_2$ and no other contrary information. The energy scale of the excitations require starting with exchange interactions of the order 10 meV. The maximum inelastic energy of 65 meV of the excitation was determined from cuts to the data. This energy was reproduced within the model by a suitable $J$ value to provide a constraint. The spin-gap is well-defined and could be constrained to a high degree of accuracy using the instrument resolution, fits to the data are shown in Fig.~\ref{Fig3}(b) to 3.9(1) meV. To model the full spectrum the higher energy scattering was artificially broadened in energy and momentum to account for the damping from Fe(2) vacancies. Both the energy and intensity within neutron scattering carries the requisite information and allows a constraint of the model. Fits to the data for low energy cuts are shown in Fig.~\ref{Fig4}(a) and a comparison of the data and model for slices of the low energy and full dispersion from BZ center is shown in Fig.~\ref{Fig4}(b)-(c). To further compare the model and data a constant energy slice in the (H,K) plane is presented in Fig.~\ref{Fig4}(d). For all calculations shown in Fig.~\ref{Fig3} and Fig.~\ref{Fig4} the data was suitably modeled using $J_1$=$J_2$=11 meV and a single-ion anisotropy of 0.95 meV. These values are close to those predicted from previous DFT studies \cite{GateTuneableFGT} and should provide a useful bases for future theoretical studies into FGT.

\begin{figure}[tb]
	\centering         
	\includegraphics[trim=0cm 13.5cm 0cm 0.5cm,clip=true, width=0.9\columnwidth]{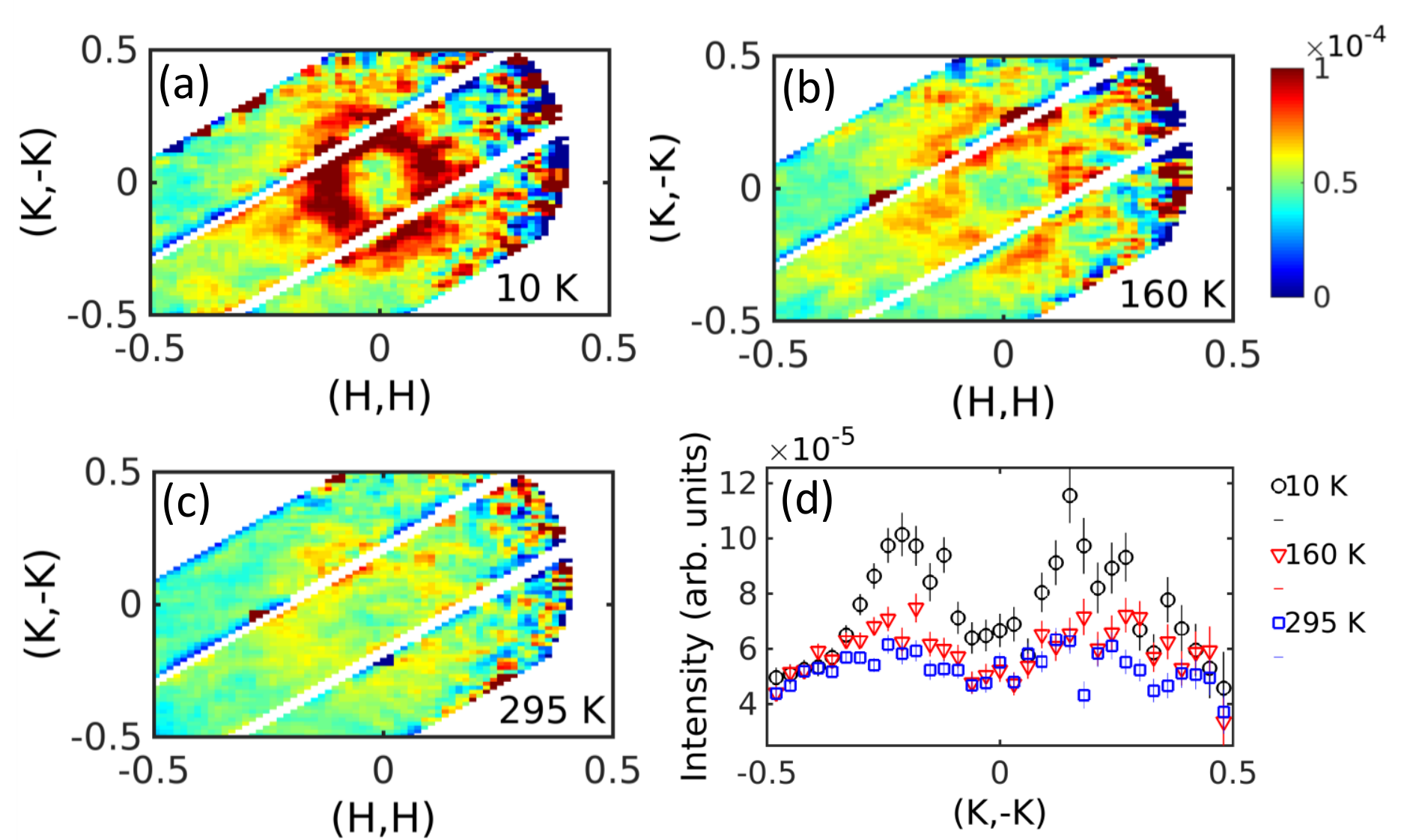}           
	\caption{\label{Fig5} Temperature dependence of inelastic scattering centered at 15 meV in the range 13.5 meV to 16.5 meV for (a) 10 K, (b) 160 K and (c) 295 K. (d) Cut through (H,H)=0 in the range -0.05 to 0.05 r.l.u. and over the L range -3 to 3 r.l.u. All the data shown have been corrected with the Bose factor. Incident energy was E$\rm _i$=30meV.}
\end{figure} 

Temperature dependence of the inelastic scattering was measured and results shown in Fig.~\ref{Fig5} at E=15 meV. The temperatures chosen correspond to the fully ordered regime (10 K), just above T$\rm _c$ (160 K) and well above T$\rm _c$ (300 K). The ring of scattering in the HK plane that is present within the magnetically ordered regime is seen to significantly decrease in intensity with increasing temperature, as expected for a magnetic excitation. However, it is still present at 160 K and at a further reduced intensity at 300 K. Correlations above T$\rm _c$ are often observed, particularly in low dimensional systems where the exchange interactions in the plane are much larger than the out of plane interactions and so persist above the T$\rm _c$ energy scale. However the presence at 300 K, around twice T$\rm _c$, is not typical and would suggest strong 2D correlations. Therefore, care must be taken in this interpretation since phonon scattering at 295 K may become an important factor, even at low Q and after Bose correction.

Collectively  the INS investigation presented has allowed access to the magnetic Hamiltonian of FGT. This is applicable for the bulk compound, however the 2D nature of the excitations allow for extension of the applicability to reduced layered and even monolayered FGT. This study was performed on Fe$_{2.75}$GeTe$_2$. The Fe site occupancy can act as a control of T$\rm _c$ and alter the exchange values, however it introduces damping of the excitations that preclude detailed modeling. From magnetization measurements the anisotropy and $\rm T_c$ both decrease as the Fe(2) occupation decreases. Testing of these altered  parameters in the Hamiltonian and extracting further nearest neighbor interactions would require the growth of larger single crystals of FGT and therefore remains as a future endeavor.



\section{\label{sec:Conclusion}Conclusion}

The magnetic correlations in Fe$_{2.75}$GeTe$_2$ have been investigated with INS on a gram sized array of single crystals. The results show a well-defined spin-gap at magnetic zone center of 3.9 meV. As the excitations disperse to higher energies they become significantly damped, due to a combination of disorder from the reduced Fe(2) site occupancy and the itinerant nature of the magnetism. To extract the relevant energy scales of the magnetic exchange interactions a minimal model consisting of nearest neighbor interactions is applied with interactions of 11 meV. The magnetic correlations show negligible inter-layer magnetic interactions, consistent with the 2D layered structure, indicating the bulk system behaves analogously to the single-layer. 

\begin{acknowledgments}
We greatfully acknowledge A.~D.~Christianson for support with the ARCS measurements and T.~Berlijn and R.~S.~Fishman for useful discussions. This research used resources at the High Flux Isotope Reactor and Spallation Neutron Source, a DOE Office of Science User Facility operated by the Oak Ridge National Laboratory. AFM (synthesis and characterization) was supported by the U. S. Department of Energy, Office of Science, Basic Energy Sciences, Materials Sciences and Engineering Division.
\end{acknowledgments}


%

\end{document}